 \documentclass[usenatbib,usedcolumn]{mnras}
\usepackage{mathptmx}

\usepackage[T1]{fontenc}
\usepackage{ae,aecompl}

\usepackage{graphicx}	
\usepackage{amsmath}	
\usepackage{amssymb}	

\title[Magnetic reconnection and Blandford-Znajek process]{Magnetic reconnection and Blandford-Znajek process around rotating black holes}

\author[C. B. Singh et al.]{
Chandra B. Singh,$^{1,2,4}$\thanks{E-mail: chandratalk@gmail.com (CBS)}
David Garofalo,$^{3,4}$\thanks{E-mail : dgarofal@kennesaw.edu (DG)}
and Elisabete M. de Gouveia Dal Pino$^{2}$
\\
$^{1}$The Raymond and Beverly Sackler School of Physics and Astronomy, Tel Aviv University, Tel Aviv 69978, Israel\\
$^{2}$Department of Astronomy (IAG-USP), University of Sao Paulo, Sao Paulo, Brazil\\
$^{3}$Department of Physics, Kennesaw State University, Marietta GA 30060, USA\\
$^{4}$Equal first authors 
}

\date{Accepted XXX. Received YYY; in original form ZZZ}

\pubyear{2018}

\begin{document}
\label{firstpage}
\pagerange{\pageref{firstpage}--\pageref{lastpage}}
\maketitle

\begin{abstract}
We provide a semi-analytic comparison between the Blandford-Znajek (BZ) and the magnetic reconnection power for accreting black holes in
the curved spacetime of a rotating black hole. Our main result is that for a realistic range of astrophysical parameters, the reconnection
power may compete with the BZ power. The field lines anchored close to or on the black hole usually evolve to open field lines in general
relativistic magnetohydrodynamic (GRMHD) simulations. The BZ power is dependent on the black hole spin while magnetic reconnection power
is independent of it for the force-free magnetic configuration with open field lines adopted in our theoretical study.
This has obvious consequences for the time evolution of such systems particularly in the context of black hole X-ray binary state 
transitions. Our results provide analytical justification of the results obtained in GRMHD simulations.
\end{abstract}




\section{Introduction}
Outflows and jets are ubiquitous in various astrophysical sources ranging from young stellar objects to black holes. Usually
relativistic jets are observed in X-ray binaries (XRBs), active galactic nuclei (AGN) and gamma-ray bursts (GRBs) which are
likely to have black holes as central compact objects \citep[]{bbr84}. There are likely to be different
mechanisms working in connection with origin, acceleration and collimation of jets, namely, thermal pressure, radiation pressure
and magnetohydrodynamics (MHD) processes. These mechanisms may be either at work in different sources or in the same source at 
different scales  (e.g., \citealt{bes10}). Based on the role of magnetic fields, there are two popular models: the jet is
powered by accretion via a magneto-centrifugal mechanism known as the Blandford-Payne process \citep{bp82} 
and the Blandford-Znajek mechanism (\citealt{bz77}; \citealt{lee00} and references therein) wherein the rotation of black hole
powers the jet through the magnetic lines anchored in its horizon. During the accretion process, large scale magnetic field
lines are likely to be dragged along with the accreting matter and field lines may accumulate and become dynamically important
with the ram pressure of the accretion flow balanced by magnetic pressure, a scenario referred to as Magnetically
Arrested Disk (MAD)(\citealt{nar03}; \citealt{br74}). Pseudo-Newtonian \citep{igu08} as well as general relativistic
(GR) MHD (\citealt{tmn11} ; \citealt {mtb12}) simulations have confirmed such highly efficient accretion flow. The topology
of field lines can also play important roles: quadrupole loops, for instance, can lead to less powerful and episodic jets 
compared to dipolar ones \citep{bec08}.

Along with these processes, magnetic reconnection may also play a role in extracting energy efficiently from the black hole
surroundings (\citealt {dl05}; \citealt{dpk10}; \citealt{kds15}; \citealt{sdk15}) and in powering jets 
(e.g., \citealt{gb09}; \citealt{zy11}; \citealt{smd16}). 
Fast magnetic reconnection can occur independently of microscopic plasma properties in presence of turbulence 
\citep{lv99} and has been tested numerically in three dimensional (3D) MHD Newtonian \citep{ketal09, ketal12} as well as
 special relativistic MHD regimes \citep{tak15}. Even from initial weak noise of velocity
 fluctuations, the process of reconnection can generate turbulence which in turn may drive it faster ( \citealt{ber17};
\citealt{ketal17}). First-order Fermi acceleration can occur in magnetically dominated environments due to fast magnetic reconnection 
\citep{dl05, dru12} and this process has been also tested successfully both via MHD
 ( \citealt{ketal11}; \citealt{ketal12}; \citealt{del16}; \citealt{bli16}) and kinetic PIC
simulations (e.g., \citealt{dra06}; \citealt{zh08}; \citealt{ss14}; \citealt{guoetal14}; \citealt{wu16}; \citealt{wu17}).

Frame dragging effects of rotating black holes can lead to complicated magnetic current sheet structures and occurrence of
magnetic reconnection can accelerate plasma \citep{ka08, kkk12}. Pseudo-Newtonian \citep{mm03, igu09}, force-free \citep{par15} and
 GRMHD simulations \citep{hietal04, ball16, ball18, dex14} have confirmed the polarity inversion of field lines and formation of magnetic
reconnection sites around spinning black holes. The regions of high current density are possible locations of high magnetic
dissipation hence the reconnection sites and can lead to acceleration of non-thermal electrons and X-ray and gamma-ray flaring.
Taking into account small scale magnetic flux loops in presence of turbulence, axisymmetric force free simulations have
shown that relativistic jets can be driven by magnetic reconnection and may be responsible for hard X-ray emission
(e.g. \citealt{par15}), as predicted in \citet{dl05} using a Newtonian approach. In another work dealing with
large scale magnetic fields, \citet{sem14} showed that reconnection cuts flux tubes and causes the ejection of
plasmoids that mimic a particle-like Penrose mechanism as it carries extracted black hole rotational energy. Long term
evolution of GRMHD simulations of MAD or magnetically choked accretion flow (MCAF) by \citet{mtb12}
 for black hole of spin 0.9375 and poloidal field with flipping type showed that reconnection of field lines in the inner
region of accretion flows around black holes can drive a transient outflow which is as powerful as a steady BZ jet \citep{dex14}.
 This can be associated with change of states in black hole X-ray binaries. Further continuing the same set of GRMHD
simulations, \citet{rpm16} showed that for a population of thermal electrons, during launching of the
transient plasmoid, when BZ jet is quenched, the ratio of gamma-ray luminosity to X-ray luminosity becomes less than 1
while in the case of the steady BZ jet it remains greater than 1.

\citet{bz13}, using a force free analytical solution under the assumption of axisymmetric and stationary
flow, could explain the profile of the angular velocity of the field lines which were obtained from the 3D GRMHD simulations of
\citet{mtb12}. Motivated by the success of this analytical study to explain the numerical results, we
here apply fast reconnection theory to address the properties that are seen in the simulations of transient jets with respect to
the black hole spin. Our goal is to analytically explore the contribution of curved spacetime and black hole spin to fast 
magnetic reconnection in order to try to shed light on some of the results seen in numerical simulations of black hole accretion. To
accomplish this we develop a solution that to a first approximation is removed from force-free magnetic field configuration.
In section~\ref{sec:2} we present the equations and expressions relevant to build a force-free black hole
magnetosphere. In section~\ref{sec:3} the magnetic reconnection power and BZ power are discussed depending upon various accretion
parameters and black hole properties. Section~\ref{sec:4} deals with summary and conclusions of our work.

\section{A force-free Kerr black hole magnetosphere}
\label{sec:2}
In this section we develop a general relativistic expression for the magnetic reconnection power from a rotating black hole. Our
background spacetime is that of Kerr which we describe using Boyer-Lindquist coordinates ($t$, $r$, $\theta$, $\phi$) 
\footnote {We note that the polar coordinate $\theta$ varies between 0 (on the rotation axis) and $\pi/2$ (on the equator), and 
the azimuthal coordinate $\phi$ between 0 and $2\pi$.} which are well known to be singular at the null horizon:

\begin{equation}
\begin{split}
 dS^{2} = -(1 - 2Mr/\rho^{2})dt^{2} - (4 M a r sin^{2}\theta/\rho^{2}) dt d\phi + (\Sigma^{2}/\rho^{2})sin^{2}\theta d\phi^{2} \\
 + (\rho^2/\Delta)dr^{2} + \rho^{2}d\theta^{2},      
  \label{eq:1}
\end{split} 
\end{equation}

with $\rho^{2} = r^{2} + a^{2}r_{g}^{2}cos^2\theta$, $\Delta = r^{2} - 2r_{g}r + r_{g}^{2}a^{2}$ and 
$\Sigma^{2} = (r^{2} + a^{2}r_{g}^{2})^{2} - a^2 r_{g}^{2} \Delta sin^2\theta$ 
where $a$, $M$ and $r_{g} = GM/c^{2}$ are the angular momentum per unit mass, mass and gravitational radius of the black hole 
respectively. Here $G$ and $c$ are the gravitational constant and speed of light, respectively.
The associated contravariant metric tensors are given as follows : $g^{tt} = - \frac{\Sigma^{2} c^{2}}{\rho^{2} \Delta}$, 
$g^{rr} = \frac{\Delta}{\rho^{2}}$, $g^{\theta \theta} = \frac{1}{\rho^{2}}$, $g^{\phi \phi} = \frac{\Delta - a^{2}r_{g}^{2}sin^2\theta}
{\rho^{2} \Delta sin^2\theta}$ and $g^{t\phi} = -\frac{2arr_{g}}{\rho^{2}\Delta}$ \citep{chan83}.\\

The singular nature of the horizon is not a problem for our analysis since we are interested in exploring a region of spacetime
outside both the black hole and the accretion disk, a hypersurface in $r$, $\theta$ spanning the radial location $r$ of 10 to 50$r_{g}$,
and a coordinate $\theta$ range of 30$^\circ$ to 45 $^\circ$. We let $a$
vary across the entire possible range. In addition to the assumption of a background Kerr spacetime, we will assume that the
magnetosphere is to first order force-free. Although such a configuration is by construction non-dissipative, we will explore
the possible consequences of a magnetosphere that evolves away from force-freeness and how this evolution affects reconnection.
This force-free approach is taken because of the simplicity introduced in the system of equations, which allows for the development
of a family of analytic solutions for the magnetic flux around the black hole. Besides, a force- free configuration is naturally
expected in the coronal regions around black-hole accretion disk system (e.g. \citealt{hietal04}). The equations that govern the 
system are as follows. The force-free condition is

\begin{equation}
F_{ab}j^{b} = 0,      
  \label{eq:2}
\end{equation}

where $F_{ab}$ is the Faraday tensor and $j^{b}$ is the current 4-vector. In terms of the vector potential we have

\begin{equation}
F_{ab} = A_{b,a} - A_{a,b}.     
  \label{eq:3}
\end{equation}

As in \citet{bz77}, the force-free condition allows to define the following function $\omega$ as

\begin{equation}
\omega = -\frac{A_{t, r}}{A_{\phi, r}} = \frac{A_{t, \theta}}{A_{\phi, \theta}},      
  \label{eq:4}
\end{equation}

where the right hand side involves radial and angular partial derivatives of the vector potential one-form components. We assume
a steady-state configuration and axisymmetry, which specifies the gauge. The importance of the force-free assumption in developing
an analytic solution is due to the function in equation ~(\ref{eq:4}), which can be thought of as the angular velocity of the
field lines for which there is physical intuition that we can appeal to. We will determine this function
by assuming that the magnetic flux is frozen into the accretion disk plasma in the equatorial plane so that $\omega$ reduces
to the Keplerian function at $\theta = 90 {^\circ}$. In addition, we assume there is a tenuous plasma in order to define a velocity
field and impose the zero proper electric field condition

\begin{equation}
F_{ab} U^{b}= 0,      
  \label{eq:5}
\end{equation}

where $U^{b}$ is the 4-velocity and components are taken from equation (13) of \citet{yiy91} as follows:

\begin{equation}
U^{t} = \frac{\Sigma^{2}}{\rho^{2} \Delta},\quad U^{r} = -\frac {[2r_{g}r(r^{2} + a^{2}r_{g}^{2})]^{1/2}}{\rho^{2}},
        \quad U^{\theta} = 0, \quad U^{\phi} = \frac{2ar_{g}r}{\rho^{2}\Delta}.      
  \label{eq:6}
\end{equation}

The invariant magnetic flux, $A_{\phi}$ has been prescribed as \citep{bz77}

\begin{equation}
A_{\phi} = br(1 - cos\theta).      
  \label{eq:7}
\end{equation}

This gives the magnetic flux as a function of $r$ and $\theta$. It is a typical solution for paraboloidal magnetic field configuration and seen in GRMHD simulations as well (e.g., \citealt{mn07}). 

Since our solution corresponds to force-free regime around a black hole, the Alfven speed is of the order of the light speed $c$ 
(e.g. \citealt{dl05}, \citealt{rpm16}). Our primary goal will be to anchor the conversation in a general relativistic analog of
magnetic energy density. For that purpose we evaluate the invariant $F_{ab}F^{ab}$ which is obtained as follows 

\begin{equation}
F_{ab}F^{ab} = 2 \alpha B_{r}^{2}g^{\theta \theta} +2 \beta B_{\theta}^{2}g^{rr}+2 \gamma B_{\phi}^{2}+4 \delta B_{\theta}B_{\phi}g^{rr}+4\xi B_{r} B_{\phi}g^{\theta \theta},
 \label{eq:8}
\end{equation}

where $B$'s are the  magnetic field components using the vector potential given in equation~(\ref{eq:7}) and 
can be expressed as $B_{r} = -brsin\theta$ and $B_{\theta} = b(1-cos\theta)$. Here,\\~\\ 
$\alpha = g^{\phi \phi}-(2g^{t \phi}-g^{tt})(U^{\phi}/U^{t})^{2}$,\\~\\ 
$\beta = g^{\phi \phi}-2g^{t \phi}(U^{\phi}/U^{t}) + g^{tt}(U^{\phi}/U^{t})^{2}$,\\~\\
$\gamma = g^{rr}g^{\theta \theta}+g^{tt}g^{rr}(U^{\theta}/U^{t})^{2}+g^{tt}g^{\theta \theta}(U^{r}/U^{t})^{2}$, \\~\\ 
$\delta = g^{t\phi}(U^{\theta}/U^{t})-g^{tt}[U^{\theta}U^{\phi}/(U^{t})^{2}]$,\\~\\ 
$\xi = g^{t \phi}(U^{r}/U^{t})-g^{tt}[U^{r}U^{\phi}/(U^{t})^{2}]$. \\~\\

Note that the magnetic field components in the Boyer Lindquist frame ($B^{a}$) are obtained from $g^{ac}g^{bd}F_{ab}$.
Therefore, the background Kerr metric appears in physical magnetic field components as well as in the magnetic energy density $F_{ab}F^{ab}$.
It is important to emphasize that $F_{ab}F^{ab}$ is therefore expressed fully in the background Kerr metric. It is the contribution of that parameter 
for magnetic reconnection that we are most interested in and has been used in equation~(\ref{eq:11}) as well as equation~(\ref{eq:12}) .

From the simulations we know that our prespription is approximately force-free in the coronal region close to funnel 
wall region like around $30{^\circ}$ from the rotation axis such that beyond a certain threshold value of $45{^\circ}$,
 the solution is no longer force-free. As a result, we limit our analytic explorations to regions in the coronal region 
with angles smaller than $45{^\circ}$. The disk region is thus outside the scope of our work.\\

In order to determine the coefficient $b$ appearing in equation~(\ref{eq:7}), we consider the equilibrium condition 
between ram pressure of freely-falling matter and magnetic pressure on the equatorial plane $\theta = 90 {^\circ}$ and at 10$r_{g}$ for non-rotating
black hole ($a = 0$) which can be expressed as 

\begin{equation}
 \frac {F_{ab}F^{ab}}{16\pi} = \frac{GM}{r^{2}} \frac{\dot M}{2 \pi r v_{in}}. 
  \label{eq:9}
\end{equation}

Here, $v_{in}$ being the infall velocity of the matter which can be certain fraction of free-fall velocity $v_{ff} = -\sqrt {2GM/r}$. The corresponding 
four velocity components are given by $(U^{t})_{eq} = [r/(r - 2r_{g})]$, $(U^{r})_{eq} = \sqrt{2r_{g}/r}$, 
$(U^{\theta})_{eq} = (U^{\phi})_{eq} = 0$. The magnetic field components are expressed as $(B_{r})_{eq} = -br$ and 
$(B_{\theta})_{eq} = b$. The contravariant metric tensors become $(g^{tt})_{eq} = -(1 - \frac{2r_{g}}{r})c^{2}$,
$(g^{rr})_{eq} = 1 - \frac{2r_{g}}{r}$, $(g^{\theta \theta})_{eq} = (g^{\phi \phi})_{eq} = \frac{1}{r^{2}}$ and 
$(g^{t \phi})_{eq} = 0$. The subscript $eq$ indicates the corresponding expressions for the accretion flow in the equatorial plane.
As a result of our calculations,

\begin{equation}
b = \left[\frac{4GM\dot M}{v_{in}(r - r_{g})[2+r^{2}-\frac{2c^{2}r_{g}}{r}(1+r^{2})(r-2r_{g})^{2}]}\right]^{1/2}.
 \label{eq:10}
\end{equation}

Following expressions in \citet{sdk15} the magnetic reconnection power release from magnetic discontinuity in the freely
falling plasma around the black hole is

\begin{equation}
P_{MB} = \frac{F_{ab}F^{ab}}{16\pi} v_{rec} 4 \pi L_{X} R_{X},     
  \label{eq:11}
\end{equation}

where $v_{rec}$ is the reconnection velocity and is typically a few percent of the Alfven speed \citep{lv99, tak15, ber17, ketal17},
and $L_{X}$ and $R_{X}$ are the reconnection region length and the inner radius location of the current sheet, respectively.
For our comparative study, BZ power is given by \citep{meier12}

\begin{equation}
P_{BZ} = \frac {F_{ab}F^{ab}}{32} r^{2}_{g} c a^{2} .     
  \label{eq:12}
\end{equation}

\section{Results}
\label{sec:3}
In this section we evaluate the dependence of the reconnection power on $r$, $\theta$, $a$, mass accretion rate $\dot M$ and
$M$. We then determine the parameters that minimize and maximize the reconnection power and compare it to the Blandford- Znajek
power extracted from black hole rotation.

\subsection{Radial dependence}
\begin{figure}
\includegraphics[width= \columnwidth]{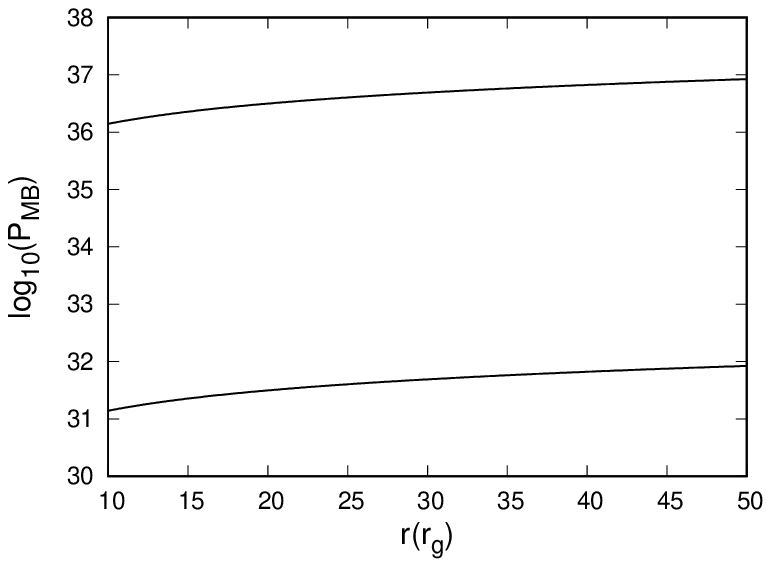}\\
\includegraphics[width= \columnwidth]{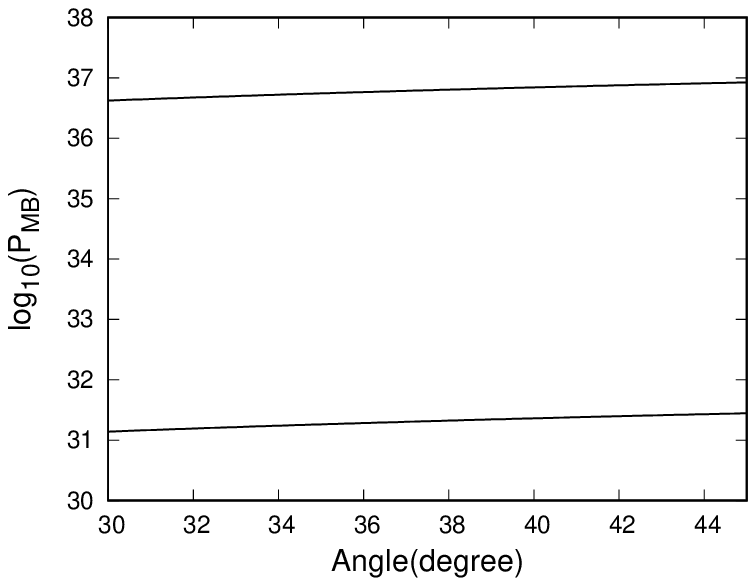}
\caption{Magnetic reconnection power $P_{MB}$ (in units of erg/s) versus radial location $r$ (in units of gravitational radius, $r_{g}$)
(top panel), and versus angle $\theta$ (in units of degree) (bottom panel). The curves give upper and lower bounds for the
$P_{MB}$ calculated for a parametric space (see text for details).}
\label{fig:1}
\end{figure}

We begin by evaluating equation~(\ref{eq:11}) as a function of $r$. For this initial analysis, the parameters are fixed in the
following way: we adopt a characteristic value of $a$ = 0.9375 and $M$ = 10$M_\odot$(solar masses) based on the results of the
GRMHD simulations \citep{dex14}. To calculate the reconnection power, we considered polar angles in the range $\theta$ 
= $30{^\circ} - 45{^\circ}$, radial inflow velocities ($v_{in}$) as 1{\%} - 10{\%} of the free-fall velocity $v_{ff}$ 
\citep{nar03}, mass accretion rates $\dot {M}$ in the range of 0.05 to 0.0005 of the Eddington value and the length of the current
sheet $L_{X}$ between 0.1 and 1 times the radial location $R_{X}$. The reconnection velocity
$v_{rec}$ = 1{\%} - 5{\%} of $c$. Top panel of Fig.~\ref{fig:1} shows the upper and lower bounds of the reconnection power for
this parametric space. The upper curve corresponds to the maximum values of the parameters above, while the lower curve
corresponds to the minimum values (except $v_{in}$, for which the lower value corresponds to the upper curve). 
The reconnection power in both curves is smaller at 10 gravitational radii ($r_{g}$) and increases outward by about order of unity at 50$r_{g}$.

\subsection{Angular dependence}
In the bottom panel of  Fig.~\ref{fig:1} we report the results of a similar analysis for the reconnection power, obtained by 
fixing $r$ at 50 $r_{g}$, and adopting the values for the remaining parameters that maximized the power in the upper panel,
namely a 1{\%} of $v_{ff}$, a 0.05 Eddington accretion rate, a current sheet that is equal in length to the radial location
or to its coordinate value, $v_{rec}$ as 5{\%} of $c$, and letting the polar angle $\theta$ to vary. In other words, we
explore the reconnection power along an arc. The result is displayed in the upper curve. The lower curve reflects the
angular dependence of the power for fixed radial location of 10 $r_{g}$, 0.0005 Eddington accretion rate, $v_{rec}$ as 1{\%} of
c, current sheet length $L_{X}$ about 10{\%} of the radial location value $R_{X}$ and $v_{in}$ as 10{\%} of $v_{ff}$. 
Similar to the radial dependence, the angular range introduces less than an order of magnitude difference in 
reconnection power, the greater power occurring closer to the equatorial plane. We have constrained our angular
range to be compatible with the results of numerical simulations that suggest that force-freeness does not extend to higher
angular regions \citep{rpm16}.

\subsection{Mass accretion rate dependence}
\begin{figure}
\includegraphics[width= \columnwidth]{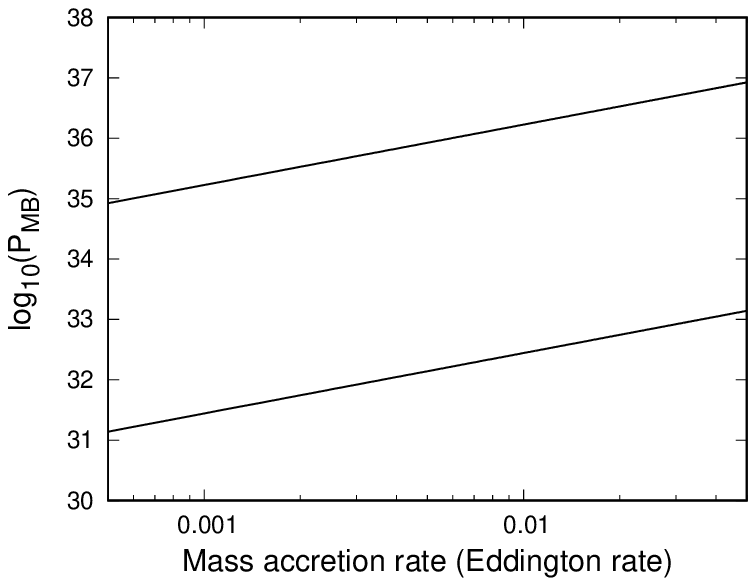}\\
\includegraphics[width= \columnwidth]{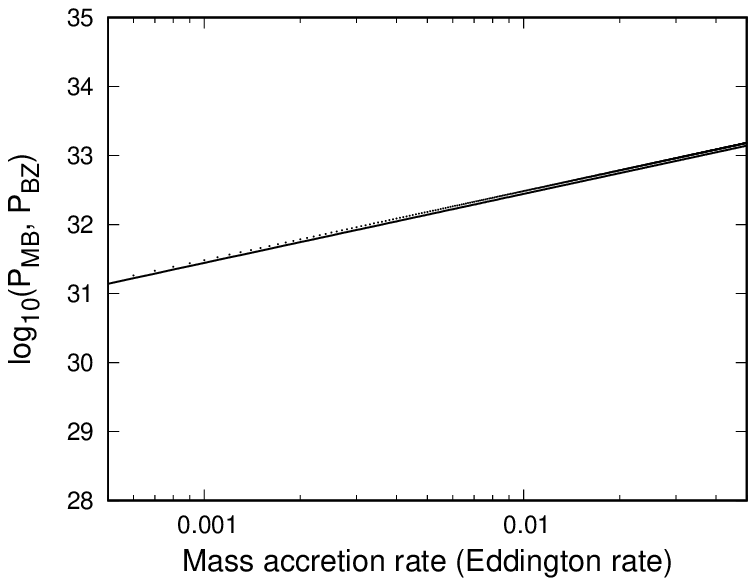}
\caption{Magnetic reconnection power $P_{MB}$ (in units of erg/s) versus mass accretion rate $\dot M$ (in units of Eddington rate) 
(top panel), and $P_{MB}$ (solid line) and Blandford-Znajek power $P_{BZ}$ (dotted line) versus mass accretion rate (in units of Eddington
rate) (bottom panel) (see text for details).}
\label{fig:2}
\end{figure}

In top panel of Fig.~\ref{fig:2}, we explore the dependence of the reconnection power on the mass accretion rate which
varies from 0.0005 to 0.05 Eddington rate for $M$ = 10 $M_{\odot}$ and $a$ = 0.9375. We find about a two order of
magnitude difference as the accretion rate
approaches the theoretical boundary between an advection dominated accretion flow and a radiatively efficient thin disk 
at about a few percent Eddington rate \citep{meier12}. The upper and lower curves are evaluated for 50 and 10$r_{g}$,
respectively, and the remaining parameters are the same as upper panel of Fig.~\ref{fig:2}.
In the bottom panel of Fig.~\ref{fig:2}, we compare the reconnection power with the Blandford-Znajek power(BZ) evaluated for 
the parameter values r = 10$r_{g}$, $\theta$ = 30, $v_{rec}$ as 1{\%} of c, current sheet length $L_{X}$ about 10{\%} of
the radial location value $R_{X}$ and $v_{in}$ as 10{\%} of $v_{ff}$.
This illustrates the important result that both powers can be comparable for reasonable parameters.

\subsection{Spin dependence}
\begin{figure}
\includegraphics[width= \columnwidth]{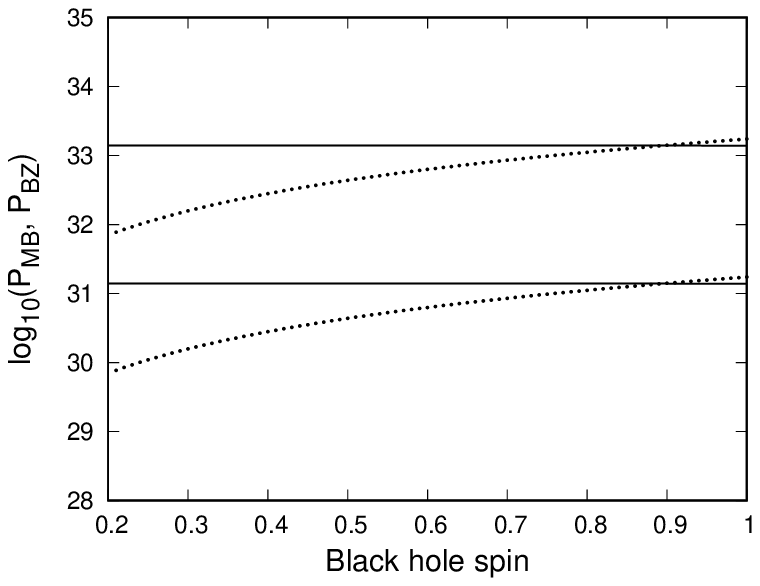}\\
\includegraphics[width= \columnwidth]{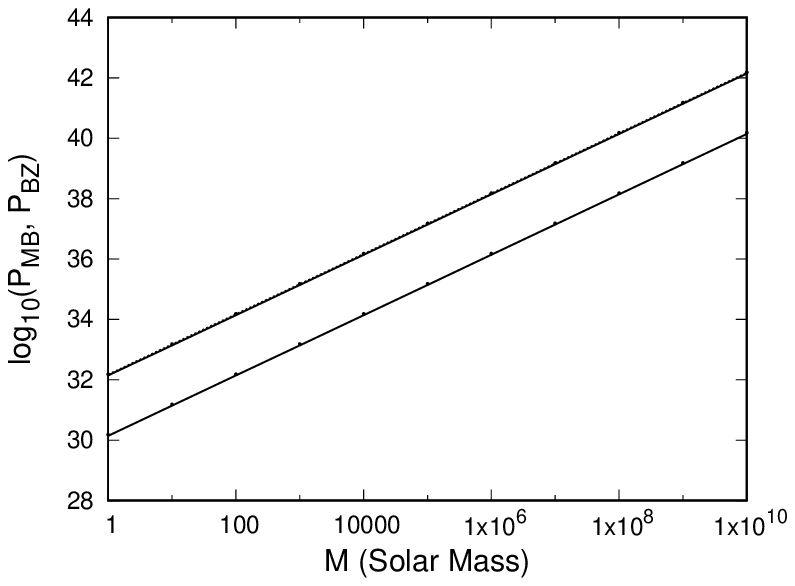}
\caption{Magnetic reconnection power $P_{MB}$ (solid lines) and Blandford-Znajek power $P_{BZ}$ (dotted lines) plotted against black hole 
spin $a$ (top panel) and black hole mass $M$ (bottom panel). The two curves coincide in the bottom panel (see text for details).}
\label{fig:3}
\end{figure}

Top panel of Fig.~\ref{fig:3} depicts the upper and lower limits for the BZ power and the reconnection power versus 
the black hole spin $a$, for the parametric space: r = 10$r_{g}$, $\theta$ = 30, $\dot M$ = 0.0005-0.05 Eddington rate,
$v_{rec}$ as 1{\%} of c, $L_{X}$ about 10{\%} of $R_{X}$, and $v_{in}$ as 10{\%} of $v_{ff}$. 
Perhaps surprisingly, we find the effect of black hole spin $a$ to be negligible for the reconnection power, while the BZ
power strongly depends on $a$, the magnetic reconnection appears to be independent of it. Usually in GRMHD simulations, field lines 
attain open configuration close to black hole (e.g. \citealt{dex14}, \citealt{rpm16}), we have also adopted similar configuration
in force-free regime for our analytical study. Our current results revealing a near independence of the magnetic reconnection power 
with black hole spin is compatible with that of GRMHD simulations \citep{dex14}.

We would like to highlight that the slopes of curves representing magnetic reconnection and BZ powers are same in case of dependence 
on mass accretion rate and black hole mass dependence due to scale invariance nature of theory however symmetry is broken for 
black hole spin.

\subsection{Mass dependence}
In bottom panel of Fig.~\ref{fig:3}, we explore the maximum and minimum reconnection power as a function of the black hole of spin 
$a$ = 0.9375 and mass which we allow to vary over 10 orders of magnitude from $1$ to $10^{10}$ solar masses, sweeping the same
 parametric space as in upper panel. This analysis shows that the conditions that maximize the reconnection power also make
the reconnection power competitive with the Blandford-Znajek power, the latter being evaluated at the same dimensionless spin
value $a$ = 0.9375 . This is not surprising as we have imposed a balance of forces in the magnetosphere which naturally scales
with the black hole mass. This result is also similar to the one found in \citep{kds15, sdk15} in the evaluation of the magnetic
reconnection power around black holes neglecting GR effects.

\subsection{Astrophysical implications}
 The takeaway message from our plots is that reconnection power can compete with the BZ power for a reasonable range
of astrophysical parameters. Since reconnection events take energy away from magnetic form, non-negligible reconnection
 makes it difficult for steady-state BZ jets. As conditions arise that make the Alfven velocity increase, the
reconnection power begins to compete with the BZ power. Magnetic reconnection partially destroys the field, but such that
the field comes back, producing the observed hysteresis of the jet and the conditions that effectively terminate the field to
cause the transition to the soft state in XRBs. In greater detail, as the accretion rate increases, both the reconnection power as well as
the magnetic field will increase so there is competition between these two effects with the accretion rate contributing to greater
BZ power but reconnection power contributing to destroying large scale field and thus weakening BZ power. On the other hand increase 
in B-field also contributes to a decrease in accretion rate so it might come natural in this overall competition to see the hysteresis
behaviour. Such time dependent phenomena need detailed study as a separate work.  

\section{Summary and conclusions}
\label{sec:4}
We have performed a general relativistic semi-analytic exploration of magnetic reconnection around rotating black holes. We
determined a range of reconnection power based on a fiducial astrophysical parameter space.
While not strictly taking into account an exact force-free magnetic configuration, we were interested
in developing an analytic solution for a black hole magnetosphere aiming at understanding where magnetic reconnection might
develop or at least constitute an effective energy release mechanism near black holes. However, since it is difficult to 
develop non force-free solutions analytically, our strategy has allowed us a complete analytic solution by considering 
force-free solution and in a way that has allowed compatibility with the numerical simulations.
The magnetic reconnection power release is minimum close to the black hole and near spin axis in the coronal region. 
It is likely to increase in strength with increase in mass accretion rate. Our main result is
that reconnection power may compete or even dominate the BZ power. This raises the question of the time evolution of such
systems and we have pointed out a reconnection-based picture of the evolution and eventual
disappearance of the BZ jet, suggesting it can provide a possible explanation for state transitions in black hole XRBs.

Following are the main highlights of our work in agreement with simulation works (e.g. \citealt{dex14}):

1. The power available from both the BZ mechanism and fast magnetic reconnection is comparable for
rapidly rotating black holes and same range of mass accretion rates.

2. The BZ mechanism is dependent on the black hole spin while the fast magnetic reconnection process in the
proximity of the black hole seems to be independent of the black hole spin.

3. The BZ process can be quenched in the presence of fast magnetic reconnection. This may occur due to
the destruction of large scale magnetic fields following magnetic reconnection events. Therefore, one can argue that the state
transitions of stellar mass black hole accretion disk systems could be triggered by this process (as suggested in e.g. \citealt
{dpk10}, \citealt{kds15}).

In brief, we mainly used the conditions available in simulation set up regarding accretion flow and black hole properties in 
the analytical theory of BZ and magnetic reconnection so that we can give theoretical interpretation of phenomena seen in 
simulation results.

\section*{Acknowledgements}
CBS acknowledges support from the Brazilian agency FAPESP (grant 2013/09065-8) and was also supported by the I-Core centre of
 excellence of the CHE-ISF. EMGDP acknowledges partial support from FAPESP (2013/10559-5) and CnPq (306598/2009-4) grants.

\bsp    
\label{lastpage}
\end{document}